\documentclass[12pt]{article}
\usepackage{amsmath,amssymb,amsfonts,amsthm}

\title{Entanglement, quantum statistics and squeezing of two $\Xi$-type three-level atoms interacting nonlinearly with a single-mode field}

\author{H R Baghshahi$^{1,2,3}$, M K Tavassoly$^{1,2,*}$ \\
 \footnotesize{$^1$ Atomic and Molecular Group, Faculty of Physics, Yazd University, Yazd, Iran} \\
 \footnotesize{$^2$ The Laboratory of Quantum Information Processing, Yazd University, Yazd, Iran} \\
 \footnotesize{$^3$ Department of Physics, Faculty of Science, Vali-e-Asr University of Rafsanjan, Rafsanjan, Iran} \\
  \footnotesize{$^*$ E-mail: mktavassoly@yazd.ac.ir}}

\begin{document}
\maketitle

 \newcommand{\norm}[1]{\left\Vert#1\right\Vert}
 \newcommand{\abs}[1]{\left\vert#1\right\vert}
 \newcommand{\set}[1]{\left\{#1\right\}}
 \newcommand{\R}{\mathbb R}
 \newcommand{\I}{\mathbb{I}}
 \newcommand{\C}{\mathbb C}
 \newcommand{\eps}{\varepsilon}
 \newcommand{\To}{\longrightarrow}
 \newcommand{\BX}{\mathbf{B}(X)}
 \newcommand{\HH}{\mathfrak{H}}
 \newcommand{\A}{\mathcal{A}}
 \newcommand{\D}{\mathcal{D}}
 \newcommand{\N}{\mathcal{N}}
 \newcommand{\x}{\mathcal{x}}
 \newcommand{\p}{\mathcal{p}}
 \newcommand{\la}{\lambda}
 \newcommand{\af}{a^{ }_F}
 \newcommand{\afd}{a^\dag_F}
 \newcommand{\afy}{a^{ }_{F^{-1}}}
 \newcommand{\afdy}{a^\dag_{F^{-1}}}
 \newcommand{\fn}{\phi^{ }_n}
 \newcommand{\HD}{\hat{\mathcal{H}}}
 \newcommand{\HDD}{\mathcal{H}}

 \begin{abstract}
The interaction between two $\Xi$-type three-level atoms and a single-mode cavity field in the intensity dependent coupling regime has been studied. Exact analytical solution of the wave function for the considered atoms-field system has been obtained by using the Laplace transform technique when the atoms are initially prepared in the excited state and the field is in a coherent state. The presented structure has the potential ability to generate various new classes of entangled states depending on the chosen nonlinearity function. Two forms of intensity-dependent coupling as well as constant coupling are considered. Some important physical properties such as quantum entanglement, quantum statistics and quadrature squeezing of the corresponding states are investigated, numerically, by which the nonclassicality features of the produced entangled state are well-established. In particular, the effect of intensity-dependent coupling on the degree of entanglement between different bipartite partitions of the system (that is, ``atom$+$atom"-field and  ``field$+$atom"-atom) using the linear entropy is investigated. At the same time, by paying attention to the negativity as a useful measure, the entanglement between the two atoms is studied in detail.
 \end{abstract}


 \section{Introduction}\label{sec-intro}
Jaynes-Cummings model (JCM) is a standard model in quantum optics which simply describes the interaction between a two-level atom with a quantized electromagnetic field \cite{Jaynes}. This model which contains dipole as well as rotating wave approximation is not only solvable, but also reveals some interesting dynamical properties of atom-field interaction such as entanglement \cite{Buzek,Faghihi2,Miry,Faghihi3}, collapse-revival phenomenon \cite{Scully}, quadrature squeezing \cite{Meystre,Arvinda}, entropy squeezing \cite{Khalil}, sub-Poissonian statistics \cite{Filipowicz}, etc. Various generalizations have been proposed to extend JCM. As some important cases one may refer to intensity-dependent coupling \cite{Recamier,Baghshahi}, multi-mode field  \cite{Orany1,Abdalla1}, multi-level atom  \cite{Cardimona,Bagheri1} and multi-atom \cite{Mahmood,Hekmatara}.

In recent decades, a lot of attention has been particularly paid to the study of  different  configurations of a three-level atom  ($\Lambda$, $V$ and $\Xi$ type)  interacting with a single- and two-mode fields with arbitrary detuning.
Due to various usefulness of different types of three-level atomic systems in quantum optics and the quantum information processing (QIP) such as electromagnetically induced transparency (EIT) \cite{Fleishhauer}, coherent population trapping (CPT) \cite{Dalton} and quantum jumps \cite{Scully}, a lot of papers concern with the  interaction  between a three-level atom and the quantized field (see for instance \cite{Faghihi,Zait1,Eied1,Obada1,Obada2,Faghihi4}).
The interaction between a $\Lambda$-type ($V$-type) three-level atom with a single-mode cavity field in a Kerr medium with intensity-dependent coupling has been studied in \cite{Faghihi} (\cite{Zait1}). The evolution of the atomic quantum entropy and the atom-field entanglement in a system of $\Xi$-configuration three-level atom interacting with a two-mode field containing additional forms of nonlinearities in the presence of intensity-dependent atom-field coupling have been studied in \cite{Eied1}. The interaction between  $\Lambda$-type ($V$-type) three-level atom and a two-mode cavity field under a multi-photon process with additional  nonlinearities similar to Ref. \cite{Eied1} has been studied in \cite{Obada1} (\cite{Obada2}).

 On the other hand, quantum entanglement is the most striking nonclassical feature of quantum mechanics. In addition to its importance in concepts and fundamentals of quantum mechanics, it plays a key role within new information technologies and in many of the interesting applications of quantum computation, quantum information, quantum teleportation, quantum cryptography, etc \cite{Benenti,Nielsen,Ekert,Bennett}. A well-known source for the production of such states is the atom-field interaction process, using different models of interactions. Once an atom and a radiation field are entangled with each other, the atom can be fully controlled by photons of the field. Many authors have found the degree of entanglement (DEM) in engineered interactions between two-, three- and four-level atoms with cavity field while different conditions have been taken into account \cite{Zhang,Abdel-Aty,Eied}.
 Even though a lot of attention has been paid to two two-level atoms interacting with different field modes \cite{Guo} (including additional interacting terms \cite{Obada.t}), as well as one three-level atom interacting with various types of fields \cite{Faghihi,Zait1}, however, to the best of our knowledge, the interaction between two three-level atoms with even a single-mode field, its exact and entire wave-vector solution and the consequence physical properties have not still been appeared in literature. This is perhaps due to the appearance of complicated coupled differential equations which seems to be hard to solve. Moreover, the entanglement dynamics of the system composed of a $\Lambda$-type and a $V$-type atom simultaneously interacting with two coupled cavities in the resonance condition has been investigated in \cite{Dao}. A scheme to generate a maximally entangled state of two three-level atoms in a nonresonant thermal cavity has been presented in \cite{Zhang2}. Altogether, it ought to be emphasised that our attack to the outlined problem is essentially different, in the sense of the used approach, its generality, exactness and solvability. Indeed, we have found the explicit form of the general state-vector corresponding to the considered atoms-field interaction by a particular method, and then some new and interesting aspects of the obtained state are evaluated.
 Meanwhile, one of the  interesting schemes in which entanglement can be created is a system containing two three-level atoms which interacts with a single-mode field.
 Thus, in this paper we motivate to consider particularly the interaction between two $\Xi$-type three-level atoms with a single-mode cavity field in the intensity-dependent coupling regime. We would like to mention that, to the best of our knowledge, even the linear form (constant coupling) of the mentioned dynamical system has not been outlined in the literature up to now. Anyway, since the behaviour of such systems depends on the initial atoms-field state, we take the atoms to be prepared in their exited states and the field is considered to be in a (standard) coherent state. At this point, it is worth noticing that, to overcome the difficulties in the coupled differential equations we will use the Laplace transform techniques. Accordingly, we successfully find the analytical expression for the wave function of the above general dynamical system, which is indeed a new class of entangled states.

  At first the role of intensity-dependent coupling regimes on the DEM between different bipartite partitions of the considered system, in particular,  ``atom$+$atom"-field and  ``field$+$atom"-atom will be investigated using the linear entropy. In addition, the entanglement between the two atoms is discussed via the evaluation of  the negativity.
 Due to the fact that, nonclassicality of radiation field plays a central role in quantum measurement and QIP, in the continuation of the paper, after quantifying the DEM of the produced atoms-field system, we investigate sub-Poissonian statistics and quadrature squeezing as two important nonclassicality features of the considered system. As will be observed, various classes of entangled states with different amounts of DEM and nonclassicality signs will be produced by employing the intensity-dependent coupling functions appropriately.

The organization of this paper is as follows: in section 2 we introduce the Hamiltonian of the model in the interaction picture and derive an exact expression for the wave function of the system under consideration. In section 3 we employ the analytical solution of the state vector of the system to investigate the time evolution of the linear entropy, negativity, Mandel parameter, mean photon number and squeezing parameters. Finally, a summary and conclusions are presented in section 4.

 \section{The model and its solution}
Let us consider two similar three-level atoms (labeled with $A_{1}$ and $A_{2}$)  with $\Xi$-configuration (has been depicted in figure 1) with exited state $|1\rangle$, ground state $|3\rangle$ and  intermediate states $|2\rangle$ with the only allowed transitions $|1\rangle\rightarrow|2\rangle$, $|2\rangle\rightarrow|3\rangle$. These two atoms interact with a single-mode cavity field. The Hamiltonian describing this system can be written as $(\hbar=1)$:
\begin{equation} \label{1}
\hat{H}=\hat{H}_{0}+\hat{H}_{1},
\end{equation}
\begin{equation} \label{2}
\hat{H}_{0}=\Omega\hat{a}^\dagger \hat{a}+\sum_{j=A_{1}, A_{2}} (\omega_{1}|1^{j}\rangle  \langle 1^{j} |+\omega_{2} |2^{j} \rangle  \langle 2^{j} |
+\omega_{3}|3^{j}\rangle  \langle 3^{j} |) ,
  \end{equation}
   \begin{eqnarray} \label{3}
   \hat{H}_{1}&=&\sum_{j=A_{1}, A_{2}} g (\hat{\sigma}_{12}^{(j)} \hat{A}+\hat{\sigma}_{21}^{(j)}\hat{A}^\dagger)
   +g (\hat{\sigma}_{23}^{(j)} \hat{A}+\hat{\sigma}_{32}^{(j)}\hat{A}^\dagger)\nonumber\\&=&\sum_{j=A_{1}, A_{2}} g (\hat{\sigma}_{12}^{(j)} \hat{a}f(\hat{n})+\hat{\sigma}_{21}^{(j)}f(\hat{n})\hat{a}^\dagger)
   +g (\hat{\sigma}_{23}^{(j)} \hat{a}f(\hat{n})+\hat{\sigma}_{32}^{(j)}f(\hat{n})\hat{a}^\dagger),
     \end{eqnarray}
  where $\hat{\sigma}_{ik}=|i\rangle \langle k |$ is the atomic raising or lowering operator, $\hat{a}$ and $\hat{a}^\dagger $ are respectively the bosonic annihilation and creation operators of the cavity field, $g$ is the atom-field coupling constant in the absence of nonlinearity and $j$ denotes the atoms $A_{1}$ and $A_{2}$. The deformed operators $\hat{A}$ and $\hat{A}^{\dagger}$ have been defined as $\hat{A}=\hat{a} f(\hat{n})$ and $\hat{A}^\dagger=f(\hat{n})\hat{a}^\dagger$ where $f(\hat{n})$ is a function of the number operator (intensity of light), a well-known operator-valued function in the nonlinear coherent state approach \cite{Manko,Vogel,Roknizadeh}. It is convenient to work in the interaction picture, in which the Hamiltonian is generally given by
\begin{equation}\label{5}
\hat{H}_{I}=e^{i\hat{H}_{0}t}\hat{H}_{1}e^{-i\hat{H}_{0}t}.
 \end{equation}\label{5}
 Using the identity, $e^{\lambda\hat{D}}\hat{B}e^{-\lambda\hat{D}}=\hat{B}+\lambda[\hat{D},\hat{B}]+\frac{\lambda^{2}}{2!}[\hat{D},[\hat{D},\hat{B}]]+....$, we finally obtain
    \begin{equation} \label{7}
    \hat{H}_{I}=\sum_{j=A_{1}, A_{2}} g(\hat{\sigma}_{12}^{(j)} \hat{A}e^{i\Delta_{1}t}+\hat{\sigma}_{21}^{(j)}\hat{A}^\dagger e ^{-i\Delta_{1}t})
    +g(\hat{\sigma}_{23}^{(j)} \hat{A}e^{i\Delta_{2}t}+\hat{\sigma}_{32}^{(j)}\hat{A}^\dagger e^{-i\Delta_{2}t}),
      \end{equation}
      where the detuning parameters are given by $\Delta_{1}=\omega_{1}-\omega_{2}-\Omega$ and $\Delta_{2}=\omega_{2}-\omega_{3}-\Omega$.

      To obtain the wave function of the outlined system, we solve the time-dependent Schr\"{o}dinger equation $i\frac{\partial}{\partial t}|\psi(t)\rangle=\hat{H}_{I}|\psi(t)\rangle$. For the assumed system, the wave function at any time $t$ can be written in the following form:
      \begin{eqnarray}\label{8}
|\psi(t)\rangle&=&\sum_{n=0}^{\infty} [ C_{1}(n,t)|1,1,n\rangle+C_{2}(n,t)(|1,2,n\rangle+|2,1,n\rangle)\nonumber\\
&+&C_{3}(n,t)(|1,3,n\rangle+|3,1,n\rangle)+C_{4}(n,t)(|2,3,n\rangle+|3,2,n\rangle)\nonumber\\&+& C_{5}(n,t)|2,2,n\rangle+ C_{6}(n,t)|3,3,n\rangle],
 \end{eqnarray}
where the coefficients $C_{i}(n,t), i=1,2,...,6$, are the unknown probability amplitudes that should be determined. We suppose that the field is initially prepared in the coherent state $|\alpha\rangle$  and the atoms enter the cavity in the exited state $|1,1\rangle$. Thus, the initial wave function is given by:
\begin{equation}\label{9}
|\psi(t=0)\rangle=\sum_{n} C_{1}(n,0)|1,1,n\rangle, \hspace{1.5cm}C_{1}(n,0)=\exp(-\frac{|\alpha|^{2}}{2}) \frac{\alpha^{n}}{\sqrt{n!}}.
 \end{equation}

 We find the equations of motion for the time dependent probability amplitudes ($C_{i}(n,t)$) by substituting (\ref{8}) in the Schr\"{o}dinger equation. This procedure arrives us at the following six coupled differential equations:
 \begin{equation}\label{10}
\frac{dC_{1}(n,t)}{dt}= -2iV_{1n}e^{i\Delta_{1}t}C_{2}(n+1,t),
\end{equation}
\begin{eqnarray}\label{11}
\frac{dC_{2}(n+1,t)}{dt}&=& -iV_{1n}e^{-i\Delta_{1}t}C_{1}(n,t) -iV_{2n}e^{i\Delta_{2}t}C_{3}(n+2,t) \nonumber\\&-&iV_{2n}e^{i\Delta_{1}t}C_{5}(n+2,t),
\end{eqnarray}
\begin{equation}\label{12}
\frac{dC_{3}(n+2,t)}{dt}= -iV_{2n}e^{-i\Delta_{2}t}C_{2}(n+1,t) -iV_{3n}e^{i\Delta_{1}t}C_{4}(n+3,t) ,
\end{equation}
\begin{eqnarray}\label{13}
\frac{dC_{4}(n+3,t)}{dt}&=& -iV_{3n}e^{-i\Delta_{1}t}C_{3}(n+2,t) -iV_{3n}e^{-i\Delta_{2}t}C_{5}(n+2,t)\nonumber\\&-&iV_{4n}e^{i\Delta_{2}t}C_{6}(n+4,t),
\end{eqnarray}
\begin{equation}\label{14}
\frac{dC_{5}(n+2,t)}{dt}= -2iV_{2n}e^{-i\Delta_{1}t}C_{2}(n+1,t) -2iV_{3n}e^{i\Delta_{2}t}C_{4}(n+3,t),
\end{equation}
\begin{equation}\label{15}
\frac{dC_{6}(n+4,t)}{dt}= -2iV_{4n}e^{-i\Delta_{2}t}C_{4}(n+3,t),
\end{equation}
where we used the abbreviations:
\begin{equation}\label{16}
V_{jn}=f(n+j)\sqrt{n+j}, \hspace{1cm}j=1,2,3,4.
 \end{equation}
Among various ways, to solve the above equations, the Laplace transformation allows one to cast the differential Eqs. (\ref{10})- (\ref{15}) into a set of algebraic equations which can be solved in a straightforward manner \cite{Wolfang}. We then perform the inverse Laplace transformation to find the probability amplitudes themselves.
We start by introducing the Laplace transform of probability amplitude $ {C}_{1}(n,t)$ as follows:
\begin{equation}\label{17,1}
\tilde {C}_{1}(n,z)= \int_{0}^{\infty} dt e^{-st} C_{1}(n,t).
 \end{equation}
Accordingly the Laplace transformed of other probability amplitudes are defined as follows
\begin{eqnarray}\label{17,2}
\tilde {C}_{2}(n+1,z)&=& \int_{0}^{\infty} dt e^{-st} e^{i \Delta_{1}t} C_{2}(n+1,t), nonumber \\
 \tilde {C}_{3}(n+2,z)&=& \int_{0}^{\infty} dt e^{-st} e^{i (\Delta_{1}+\Delta_{2})t} C_{3}(n+2,t), nonumber \\
  \tilde {C}_{4}(n+3,z)&=& \int_{0}^{\infty} dt e^{-st} e^{i (2 \Delta_{1}+\Delta_{2})t} C_{4}(n+3,t), nonumber \\
  \tilde {C}_{5}(n+2,z)&=& \int_{0}^{\infty} dt e^{-st} e^{2 i \Delta_{1} t} C_{5}(n+3,t), nonumber \\
   \tilde {C}_{6}(n+4,z)&=& \int_{0}^{\infty} dt e^{-st} e^{2 i (\Delta_{1}+\Delta_{2}) t} C_{6}(n+4,t).
\end{eqnarray}
Now, if we multiply the  Eqs. (\ref{10})- (\ref{15}), respectively by $e^{-st}$, $e^{-st} e^{i \Delta_{1}t}$, $e^{-st} e^{i (\Delta_{1}+\Delta_{2})t}$, $e^{-st} e^{i (2 \Delta_{1}+\Delta_{2})t}$, $e^{-st} e^{2 i \Delta_{1} t}$ and $e^{-st} e^{2 i (\Delta_{1}+\Delta_{2}) t}$, and then integrate over time (using the Eqs. (\ref{17,1}), (\ref{17,2}) and integration by parts), the probability amplitudes obey the following set of equations
\begin{eqnarray} \label{18}
\hspace{-1.5cm} z\tilde{C}_{1}(n,z) = C_{1,n}(0)-2iV_{1n}\tilde{C}_{2}(n+1,z), \nonumber
\end{eqnarray}
\begin{eqnarray} \label{19}
\hspace{-1.5cm} (z -  i \Delta_{1})  \tilde{C}_{2}(n+1,z) = -iV_{1n}\tilde{C}_{1}(n,z)-iV_{2n}\tilde{C}_{3}(n+2,z)-iV_{2n}\tilde{C}_{5}(n+2,z),  \nonumber
\end{eqnarray}
\begin{eqnarray} \label{20}
\hspace{-1.5cm} (z - i(\Delta_{1}+\Delta_{2}))  \tilde{C}_{3}(n+2,z) = -iV_{2n}\tilde{C}_{2}(n+1,z)-iV_{3n}\tilde{C}_{4}(n+3,z),  \nonumber
\end{eqnarray}
\begin{eqnarray} \label{21}
\hspace{-1.5cm} (z - 2i\Delta_{1}-i\Delta_{2})\tilde{C}_{4}(n+3,z) &=& - i V_{3n}\tilde{C}_{3}(n+2,z)-iV_{3n}\tilde{C}_{5}(n+2,z)  \nonumber  \\
 &-& i V_{4n}\tilde{C}_{6}(n+4,z),  \nonumber
\end{eqnarray}
\begin{eqnarray} \label{22}
\hspace{-1.5cm} (z - 2i\Delta_{1})\tilde{C}_{5}(n+2,z) = - 2 i V_{2n}\tilde{C}_{2}(n+1,z)-2iV_{3n}\tilde{C}_{4}(n+3,z),  \nonumber
\end{eqnarray}
\begin{eqnarray} \label{23}
\hspace{-1.5cm} (z - 2 i(\Delta_{1}+\Delta_{2}))\tilde{C}_{6}(n+4,z) = -2iV_{4n}\tilde{C}_{4}(n+3,z).
\end{eqnarray}
 The above algebraic equations can be solved under resonance condition. In this case, after some lengthy but straightforward calculations, we obtain the probability amplitudes (via the inverse Laplace transform techniques) as below
 \begin{eqnarray} \label{24}
C_{1}(n,t)&=&\frac{C_{1}(n,0)}{x_{2} \eta_{n}} [(x_{2n}-x_{4n})\eta_{n}+(2V_{1n}^{2}x_{2n}-\beta_{2n}^{2}x_{4n})\cos(\beta_{1n}t)\nonumber\\
&-&(2V_{1n}^{2}x_{2n}-\beta_{1n}^{2}x_{4n})\cos(\beta_{2n}t)],
\end{eqnarray}
 \begin{eqnarray} \label{25}
C_{2}(n+1,t)&=&\frac{iC_{1}(n,0)}{2\beta_{1n}\beta_{2n} \eta_{n}}
[(x_{4n}-2\beta_{1n}^{2})\beta_{2n}\sin(\beta_{1n}t)\nonumber\\&-&(x_{4n}-2\beta_{2n}^{2})\beta_{1n}\sin(\beta_{2n}t)],
\end{eqnarray}
  \begin{eqnarray} \label{26}
C_{3}(n+2,t)&=&\frac{C_{1}(n,0)}{x_{2n}\eta_{n}}
[-x_{5n}\eta_{n}-(\beta_{2n}^{2}x_{5n}-V_{1n}V_{2n}x_{2n})\cos(\beta_{1n}t)\nonumber\\&+&(\beta_{1n}^{2}x_{5n}-V_{1n}V_{2n}x_{2n})\cos(\beta_{2n}t)],
\end{eqnarray}
\begin{equation} \label{27}
C_{4}(n+3,t)=\frac{ix_{1n}C_{1}(n,0)}{2V_{4n}\eta_{n}}
[\sin(\beta_{1n}t)/\beta_{1n}-\sin(\beta_{2n}t)/\beta_{2n}],
\end{equation}
\begin{eqnarray} \label{28}
C_{5}(n+2,t)&=&\frac{2C_{1}(n,0)}{x_{2n}\eta_{n}}
[-x_{5n}\eta_{n}-(\beta_{2n}^{2}x_{5n}-V_{1n}V_{2n}x_{2n})\cos(\beta_{1n}t)\nonumber\\&+&(\beta_{1n}^{2}x_{5n}-V_{1n}V_{2n}x_{2n})\cos(\beta_{2n}t)],
\end{eqnarray}
\begin{equation} \label{29}
C_{6}(n+4,t)=\frac{x_{1n}C_{1}(n,0)}{x_{2n}\eta_{n}}
[\eta_{n}-\beta_{1n}^{2}\cos(\beta_{2n}t)+\beta_{2n}^{2}\cos(\beta_{1n}t)],
\end{equation}
where
\begin{eqnarray}\label{30}
x_{1n}&=&6V_{1n}V_{2n}V_{3n}V_{4n},\hspace{1.cm}      x_{2n}=6V_{1n}^{2}V_{3n}^{2}+4V_{1n}^{2}V_{4n}^{2}+6V_{2n}^{2}V_{4n}^{2},
\nonumber\\x_{3n}&=&2(V_{1n}^{2}+V_{4n}^{2})+3(V_{2n}^{2}+2V_{3n}^{2}),\hspace{1.cm}
x_{4n}=6V_{1n}^{2}V_{3n}^{2}+4V_{1n}^{2}V_{4n}^{2},
\nonumber\\x_{5n}&=&2V_{1n}V_{2n}V_{4n}^{2}, \hspace{1.cm} \eta_{n}=\sqrt{x_{3n}^{2}-4x_{2n}},
\nonumber\\ \beta_{1n}&=&\sqrt{\frac{x_{3n}+\eta_{n}}{2}}, \hspace{1.cm}\beta_{2n}=\sqrt{\frac{x_{3n}-\eta_{n}}{2}}.
\end{eqnarray}
 Thus, the explicit form of the interacting field with the two three-level atoms, which from the quantum mechanical point of view contains all information about the considered system, has been obtained in the resonance condition. We end this section with mentioning another perspective of the JCM. Generally, this model and its extensions have been solved in the resonant \cite{Kayham,Basskirov,Abdalla}, off resonant \cite{Faghihi2}  and far from resonant \cite{Wolfang} cases. The entropy squeezing of a JC with Gluber-Laches state in the resonant case has been studied in \cite{Kayham}. The two-atom two-photon JCM  with resonant condition has been investigated in \cite{Basskirov}. The interaction between a two-level atom with two-mode cavity field via time-dependent coupling in the resonant case has been presented in \cite{Abdalla}, where the field entropy and the DEM have been quantified.
 %
 \section{Physical properties}
Now, which we successfully obtained the probability amplitudes (and so the explicit form of the wave function of the entire two three-level atomic system which interacts with a single mode coherent field), we are able to study the quantum dynamical properties of the atoms and field such as linear entropy, negativity, quantum statistical properties and quadratures squeezing. It is worth noticing that choosing different nonlinearity functions $f(n)$ leads to various atoms-field systems with different physical results. In this respect, we choose two nonlinearity functions in the following forms:

\textsl{(1) Center-of-mass motion of a trapped ion.} The nonlinearity function associated with this system reads as \cite{Mateos}:
\begin{equation} \label{31}
f_{TI}(n)=\frac{L_{n}^{1}(\eta^{2})}{(n+1)L_{n}^{0}(\eta^{2})}.
\end{equation}
where $\eta$ is Lamb-Dicke parameter and $L_{n}^{m}(x)$ is the associated Laguerre polynomial. The nonlinear coherent state associated with such a nonlinearity function has been introduced by Filho and Vogel in a pioneering paper \cite{Vogel}.

\textsl{(2) Harmonious states.} The harmonious state is characterized by the nonlinearity function
\begin{equation} \label{32}
f_{HS}(n)=\frac{1}{\sqrt{n}}.
\end{equation}
This function first introduced by  Man$^{,}$ko et al \cite{Manko} in relation to the harmonious state which has been introduced by Sudarshan \cite{Sudarshan}.

 \subsection{Measurement of DEM}

Due to the apparent entanglement feature of the considered atoms-field system, it is natural to investigate the amount of this pure quantum quantity at first. For achieving to this purpose, several measures of DEM have been proposed. These contain entanglement of formation \cite{Bennett2,Wootters}, relative entropy of entanglement \cite{Vedral}, entanglement of distillation \cite{Bennett3}, linear entropy of entanglement \cite{Kim}, negativity \cite{Kzyczkowski,Vidal}, concurrence \cite{Hill} and so on. In this section we use the  ``linear entropy" and  ``negativity" to discuss the DEM for the different parts of the considered system.
  \subsubsection{Linear entropy}
 For a bipartite quantum system ($A$ and $B$), the linear entropy is defined as:
\begin{equation}\label{33}
S_{i}(t) = 1 - \mathrm{Tr}(\rho_{i}^{2}(t)), \hspace{0.5cm} i = A,B,
\end{equation}
where $\rho_{i}(t)$ denotes the reduced density operator related to subsystem $ i $. \\
 In order to evaluate this quantity for our model, let consider two different bipartite systems ``$A_1+A_2$-field" and  ``$A_{2}+$ field-$A_{1}$",  where $A_{1}$ and $A_{2}$  represent the atomic subsystems being in the $\Xi$-type three-level atom.
 It may be noted that by using the the mentioned notation, for instance the case  ``$A_1+A_2$-field", we have reduced the three-part system to the bipartite one, i.e,  ``$A_1+A_2$" and  ``field"  which their entanglement is of interest (and also for the other case).
 Anyway, the density matrix of the atoms-field system is $\rho_{_{A_{1}A_{2}-F}}(t) = |\psi(t)\rangle  \langle \psi(t) |$  with $|\psi(t)\rangle $ has been given in (\ref{8}).
For the bipartite system ``$A_{1}+A_{2}$-field", the reduced density matrix of the subsystem composed of two atoms can be obtained by tracing over the field as follows:
\begin{equation}\label{31}
\rho_{A_{1}A_{2}}(t)=\mathrm{Tr}_{F}(\rho_{_{A_{1}A_{2}-F}}(t))=
\begin{pmatrix} 
\rho_{11}&\rho_{12}&\rho_{13}&\rho_{12}&\rho_{15}&\rho_{14}&\rho_{13}&\rho_{14}&\rho_{16}
\cr\rho_{21}&\rho_{22}&\rho_{23}&\rho_{22}&\rho_{25}&\rho_{24}&\rho_{23}&\rho_{24}&\rho_{26}
\cr\rho_{31}&\rho_{32}&\rho_{33}&\rho_{32}&\rho_{35}&\rho_{34}&\rho_{33}&\rho_{34}&\rho_{36}
\cr\rho_{21}&\rho_{22}&\rho_{23}&\rho_{22}&\rho_{25}&\rho_{24}&\rho_{23}&\rho_{24}&\rho_{26}
\cr\rho_{51}&\rho_{52}&\rho_{53}&\rho_{52}&\rho_{55}&\rho_{54}&\rho_{53}&\rho_{54}&\rho_{56}
\cr\rho_{41}&\rho_{42}&\rho_{43}&\rho_{42}&\rho_{45}&\rho_{44}&\rho_{43}&\rho_{44}&\rho_{46}
\cr\rho_{31}&\rho_{32}&\rho_{33}&\rho_{32}&\rho_{35}&\rho_{34}&\rho_{33}&\rho_{34}&\rho_{36}
\cr\rho_{41}&\rho_{42}&\rho_{43}&\rho_{42}&\rho_{45}&\rho_{44}&\rho_{43}&\rho_{44}&\rho_{46}
\cr\rho_{61}&\rho_{62}&\rho_{63}&\rho_{62}&\rho_{65}&\rho_{64}&\rho_{63}&\rho_{64}&\rho_{66}.
\end{pmatrix} 
\end{equation}
The matrix elements in  (\ref{31}) at any time $t$ are given as
\begin{equation} \label{32}
\rho_{ij}(t)=\sum_{n=0}^{\infty}C_{i}(n,t)C_{j}^{*}(n,t),     \hspace{1cm}i,j=1,2,....,6.
\end{equation}
The linear entropy of the subsystem (atoms) can be defined through their respective reduced density matrix as \cite{Agarwal2}:
\begin{equation}\label{33}
S_{A_{1}A_{2}}(t)=1-\mathrm{Tr}(\rho_{A_{1}A_{2}}^{2}(t)).
\end{equation}
It reflects the DEM between  ``the two atoms" and  ``the coherent field".
In figure 2, we have plotted the linear entropy of the atoms versus scaled time $gt$. In this figure and all figures which will be presented in the continuation of the paper, frames (a), (b) and (c) concern  respectively with the $f(n)=1$ (linear case), $f_{HS}(n)=1/\sqrt{n}$ (harmonious state) and $f_{TI}(n)=L_{n}^{1}(\eta^{2})/(1+n)L_{n}^{0}(\eta^{2})$ (trapped ion state). From the figure 2(a) whereas the intensity-dependent coupling is neglected, a random behavior around the value of $0.3$  for the time evolution of the linear entropy is observed. We can see from the figure 2(b) that in the presence of intensity-dependent coupling (with $f_{HS}=1/\sqrt{n}$), linear entropy oscillates regularly between zero and its maximum value ($0.05$). In this case, while the entangling is always negligible, the field and two atoms are completely disentangled at some particular times. Also, it is clear from figure 2(c) that,  for the trapped ion state, after a short time, the entropy rapidly oscillates  around its upper value (nearly $0.7$). Finally, if one compares frames 2(b) and 2(c) with 2(a), it is understood that the density-dependent couplings, harmonious (trapped ion), reduces (increases) the maximum amount of the entanglement between two atoms and field. Therefore, in this way, by appropriately choosing the nonlinearity (intensity-dependent) function one may tune the amount of entanglement.\\
For the investigation of the DEM in second bipartite system  ``$A_{2}$+field-$A_{1}$", one can obtain the reduced density matrix of the first atom ($A_{1}$) by tracing Eq. (\ref{31}) over the second atom ($A_{2}$) as follows:
\begin{equation}\label{34}
 \rho_{A_{1}}(t)=\mathrm{Tr}_{A_{2}}(\rho_{A_{1}A_{2}}(t))=
\begin{pmatrix} 
y_{11}&y_{12}&y_{13}
\cr y_{21}&y_{22}&y_{23}
\cr y_{31}&y_{32}&y_{33},
\end{pmatrix} 
\end{equation}
where
\begin{eqnarray}\label{35}
 y_{11}&=&\rho_{11}+\rho_{22}+\rho_{33},\hspace{0.5cm}  y_{12}=\rho_{12}+\rho_{25}+\rho_{34}=y_{21}^{*},\nonumber\\
   y_{13}&=&\rho_{13}+\rho_{24}+\rho_{36}=y_{31}^{*},\hspace{0.5cm}   y_{22}=\rho_{22}+\rho_{44}+\rho_{55},\nonumber\\
   y_{23}&=&\rho_{23}+\rho_{46}+\rho_{45}=y_{32}^{*},\hspace{0.5cm}   y_{33}=\rho_{33}+\rho_{44}+\rho_{66}.\nonumber\\
\end{eqnarray}
Consequently, the linear entropy of  $A_{1}$ atom as a measure of DEM between $A_{1}$ and  ``$A_{2}$+field" can be obtained by the following form
\begin{equation}\label{33}
  S_{A_{1}}(t)=1-\mathrm{Tr}(\rho_{A_{1}}^{2}(t)).
\end{equation}
In figure 3, we have plotted the linear entropy for the first atom ($A_{1}$) with the same parameters as in figure 2. This figure shows the time evolution of DEM between  one atom and the reminder of the system. As is seen from the plot 3(a), whereas the intensity-dependent coupling is disregarded, a chaotic behaviour for the time evolution of the linear entropy is revealed. The same general behaviour can be seen in plot 3(c) for the intensity-dependent coupling using the trapped ion system with $f_{TI}(n)$  in (\ref{30}). By including the intensity-dependent coupling with $f_{HS}(n)$  in (\ref{31}) corresponding to harmonious state in plot 3(b), a regular oscillatory behaviour is observed. Also, comparing the plots 2(a), 2(b) and 2(c) indicates that the intensity-dependent coupling in the harmonious state  form (figure 3(b)) reduces the value of the time-average of DEM between one atom and the reminder of the system. Altogether, their maxima are at the same order. In addition, as is readily seen, the DEM between $A_1$ and  ``$A_2$+field" (3(b)) is much larger than  DEM between field and  ``$A_1+A_2$" (2(b)).
 \subsubsection{Negativity}
In this subsection, we are mainly interested in analyzing the effect of the intensity-dependent coupling on the entanglement dynamics between "the two atoms" that coupled to the single-mode field inside the cavity. To study the dynamics of the mentioned quantity, one must choose an appropriate entanglement measure. For the present case, the negativity is a good computable measure for the DEM between the two atoms. The concept of the negativity is related to the Peres-Horodecki condition for the separability of a state \cite{Peres,Horodecki}. They proved that a necessary condition for separability is that the matrix obtained by partial transposition of $\rho$ has only non-negative eigenvalues. However, the state is entangled if one or more of the eigenvalues of partial transposition matrix is negative. Negativity for the subsystem which contains "two atoms" is defined by the  following form \cite{Vidal}
\begin{equation}\label{331}
\mathcal{N}=\frac{||\rho_{A_{1}A_{2}}^{T_{A_{2}}}||_{1}-1}{2},
\end{equation}
where $\rho_{A_{1}A_{2}}^{T_{A_{2}}}$ is the matrix obtained by partially transposing the atomic reduced density matrix with respect to the second atom ($A_{2}$), and $||\rho_{A_{1}A_{2}}^{T_{A_{2}}}||_{1}$ shows the trace norm of the operator $\rho^{T}_{A_{2}}$. The trace norm of any operator $\hat{O}$  is defined by $||\hat{O}||_{1}=\mathrm{Tr}\sqrt{\hat{O}^{\dag} \hat{O}}$ \cite{Reed}, that is equal to the sum of the absolute values of the eigenvalues of $\hat{O}$, when $\hat{O}$ is Hermitian. The reduced density matrix for atoms ($\rho_{A_{1}A_{2}}$) has positive eigenvalues and so $\mathrm{Tr}(\rho_{A_{1}A_{2}})=1$. Also, for the partial transpose of the atomic reduced matrix we have $\mathrm{Tr}(\rho_{A_{1}A_{2}}^{T_{A_{2}}})=1$, but because of the fact that it may possess negative eigenvalues, its trace norm can be written as follows \cite{Akhtarshenas}
\begin{equation}\label{332}
||\rho_{A_{1}A_{2}}^{T_{A_{2}}}||=\sum_{i}|\mu_{i}|=\sum_{i}\mu_{i}-2\sum_{i}\mu^{neg}_{i}=1-2\sum_{i}\mu^{neg}_{i},
\end{equation}
where $\mu_{i}$ and $\mu^{neg}_{i}$ are the eigenvalues and negative eigenvalues of $\rho_{A_{1}A_{2}}^{T_{A_{2}}}$, respectively. Consequently, we need only to calculate eigenvalues of $\rho_{A_{1}A_{2}}^{T_{A_{2}}}$ to arrive at the DEM between the two atoms by using the negativity measure. \\
Figure 4 shows the evolution of this quantity in terms of the scaled time $gt$ for an initial mean number of photons fixed at $|\alpha|^{2}=10$.
The temporal behaviour of the negativity for the cases  with constant coupling (4(a)) as well as nonlinear atoms-field coupling with trapped ion nonlinearity (4(c)) show irregular oscillations. As is shown in plot 4(b), for the case of Harmonious state the DEM between the two atoms displays regular periodic oscillations.  Meanwhile, in all of the mentioned cases, the maximum value of the entanglement between the two atoms occurs in a short time passing from  the beginning of the interaction between the two atoms and the single-mode cavity field. Also, comparing the plots 4(a) and 4(c) with  4(b) shows that, while the two atoms are entangled at all times for 4(a) and 4(c), for the  Harmonious state nonlinearity the entanglement between the two atoms is disappeared in at some moments of time.         
 \subsection{Quantum statistics: Mandel parameter and mean photon number distribution}
The so-called Mandel $Q$-parameter suitably describes the violation of the photon number distribution of the state from Poissonian statistics corresponding to coherent field. This parameter for a single-mode light field has been defined as follows \cite{Mandel}
 \begin{equation}\label{34}
Q=\frac{\langle \hat{n}^{2}\rangle-\langle \hat{n}\rangle^2}{\langle \hat{n}\rangle}-1.
\end{equation}
If $-1\leq Q<0$ ($Q>0$) the field statistics is sub-Poissonian (super-Poissonian) and $Q=0$ shows the Poissonian statistics.  Using the wave function for our considered system, it is easily seen that:
\begin{eqnarray}\label{35}
\langle n\rangle&=&\sum_{n=0}^{\infty}n(|C_{1}(n,t)|^{2}+|C_{5}(n,t)|^{2}+|C_{6}(n,t)|^{2}\nonumber\\
&+&2(|C_{2}(n,t)|^{2}+|C_{3}(n,t)|^{2}+|C_{4}(n,t)|^{2})),\nonumber\\
\langle n^{2}\rangle&=&\sum_{n=0}^{\infty}n^{2}(|C_{1}(n,t)|^{2}+|C_{5}(n,t)|^{2}+|C_{6}(n,t)|^{2}\nonumber\\
&+&2(|C_{2}(n,t)|^{2}+|C_{3}(n,t)|^{2}+|C_{4}(n,t)|^{2})),
\end{eqnarray}
 where $C_{i}(n,t),i=1,....,6$ have been determined in Eqs. (\ref{24})-(\ref{29}).

 In figure 5 we plotted the time evolution of Mandel parameter versus the scaled time $gt$.  For the linear case ($f(n)=1$), this quantity varies between positive and negative values, which means that the photons display super- or sub-Poissonian statistics for different intervals of times, alternatively.  But, from figure 5(b), where the nonlinear function is harmonious, we observe that the Mandel parameter possesses a regular periodic behaviour in the negative region. So, the entire atom-field state of the system has always a sub-Poissonian statistics and so is full nonclassical state. Figure 5(c) which is plotted for the trapped ion state shows the same behaviour like the figure 5(a), qualitatively. Meanwhile, a little difference may be observed between figures them in the sense that, the time average of sub-Poissonian statistics of constant coupling case (5(a)) is  greater than the case with  trapped ion nonlinearity function (5(c)).

We end this subsection with an overview on the mean photon number distribution in an explicit manner \cite{Eied,Singh}. By using the first relation in Eq. (\ref{35}), in figure 6, we have plotted the time evolution of mean photon number versus the scaled time $gt$ for the chosen parameters similar to figures 2. We can see the typical collapse-revival phenomena  as a nonclassical sign for linear case ($f(n)=1$). Also, the behaviour of mean photon number for Harmonious (trapped ion) nonlinearity is regular (chaotic).

%
 \subsection{Quadrature squeezing}
 Squeezed light is a radiation field without a classical analogue \cite{Walls}. The usefulness of such light relates to several applications like optical communication networks \cite{Shapiro}, interferometric techniques \cite{Schumaker} and optical waveguide tap \cite{Shapiro1}. To investigate the squeezing properties of the obtained state $|\psi(t)\rangle$ in (\ref{8}), we introduce two quadrature field operators $\hat{x}=(\hat{a}+\hat{a}^{\dagger})/2$ and $\hat{y}=(\hat{a}-\hat{a}^{\dagger})/2i$. These operators satisfy the uncertainty relation $(\Delta\hat{x})^{2}(\Delta\hat{y})^{2} \geq1/16$. Consequently, a state is said to be squeezed in the variable $\hat{x}$ ($ \hat{y}$) if $(\Delta\hat{x})^{2}<1/4$ ($(\Delta\hat{y})^{2}<1/4$).
However, by defining $S_{x}=4(\Delta\hat{x})^{2}-1$ and $S_{y}=4(\Delta\hat{y})^{2}-1$
squeezing occurs in $\hat{x}$ ($\hat{y}$) component if $-1<S_{x}<0$ ($-1<S_{y}<0$).

Figure 7 describes the quadrature squeezing in position  in terms of the scaled time.  For the linear regime ($f(n)=1$) and nonlinear case (trapped ion) squeezing exists in $x$  quadrature only at the beginning of the occurrence of the  atom-field interaction.
We can see from figure 7(b) that, for the nonlinear case with $f_{HS}=1/\sqrt{n}$ the squeezing exists in $x$ quadrature at all times, with a regular behaviour. So, recalling that all considered criteria in this paper are sufficient not necessary condition for nonclassicality behaviour, one may conclude that the atom-field states system with harmonious state coupling is full nonclassical state at all times.
 However,  it is worth to mention that, the depth of squeezing (negative value of squeezing parameter)
 at the initial interaction time in figures (7(a)) and (7(c)) are  greater than figure (7(b)).
%

 \section{Summary and conclusion}
 In this paper, we have considered the nonlinear as well as linear interaction between two identical $\Xi$-type three-level atoms and a single-mode field using the generalized JCM with intensity-dependent coupling between atom and field. Next, after finding the explicit form of the state vector of  the considered atoms-field system  by using the Laplace transformation techniques in a general manner, entanglement degree between different parts of the subsystems, sub-Poissonian statistics and quadrature squeezing of the obtained state have been investigated, numerically. We would like to emphasize the generality of our obtained formalism in the sense that, it may be used for any physical  system and nonlinear oscillator algebra with arbitrary function $f(n)$.  Even though the proposed structure can work with arbitrary nonlinearity function, we studied the effect of intensity-dependent coupling by considering the nonlinearity functions $f_{HS}=1/\sqrt{n}$ and $f_{TI}=L_{n}^{1}(\eta^{2})/(1+n)L_{n}^{0}(\eta^{2})$, in addition to the case with constant coupling. Briefly, the main results of the paper are listed as follows.
\begin{itemize}
  \item The intensity-dependent coupling has crucial effect on the DEM, sub-Poissonian statistics and quadrature squeezing.

    \item  The intensity-dependent coupling in the form of $f_{HS}=1/\sqrt{n}$ reduces the maximum amount of DEM between atoms and field, while for the trapped ion case it increases this value up to $0.7$, as compared with constant coupling (figure 2). While the maximum value between one atom and the reminder of the system for different used  atoms-field couplings are almost the same, for the Harmonious state nonlinearity this measure is fully destroyed at some specific moments of time (figure 3). In particular our numerical calculations for the negativity measure (showing the entanglement between the ``two atoms") associated with  the considered states of the system show that,   for the constant coupling (4(a)) as well as the trapped ion nonlinearity (4(c))  it always get a noticeable value, while for the Harmonious state it gets zero value at some moments of time (4(b)).

     \item The temporal behavior of different entanglement measures, Mandel parameter and quadrature squeezing in the presence of  intensity-dependent coupling in the form of $f_{HS}=1/\sqrt{n}$ oscillate regularly with time. The Mandel and quadrature squeezing parameters for this special case always remain in the negative region (figures 5 and 7) and mean photon number has oscillatory behaviour for all considered cases (figure 6) which means that, the corresponding atoms-field states of the system possess nonclassical features of interest.

\item
  It ought to be mentioned that, this study can be carried out by considering different configurations of  three-level atoms as well as different nonlinearity functions and we hope to report this works in the near features elsewhere.
\end{itemize}

{\bf Acknowledgement:} The authors would like to thank Dr M J Faghihi for his
                       useful discussion.


 \end{document}